\documentclass[prl,twocolumn,preprintnumbers,amsmath,amssymb]{revtex4}

\usepackage{graphicx}
\usepackage{bbm}
\usepackage{float}

\begin{document}

\preprint{}
\title{Response to ``Comment on Ferroelectrically induced weak ferromagnetism by design,'' by R. de Sousa and J. E. Moore, arXiv:0806.2142}
\author{Craig J. Fennie}
\affiliation{School of Applied and Engineering Physics, Cornell University, Ithaca, NY 14853}

\begin{abstract}
\end{abstract}


\maketitle


In Ref.~\onlinecite{fennie.prl.08} the current author  presented 
a physically intuitive set of design criteria in order to identify or
design a material in which a ferroelectric (FE) lattice distortion 
induces weak ferromagetism (wFM). The criteria target 
an antiferromagnetic-{\bf paraelectric} (AFM-PE) structure for 
which wFM is symmetry-forbidden but symmetry allowed in the FE phase.
Said another way, a FE distortion can induce wFM when the 
phenomenological invariant 
$E_{{\rm PLM}}$$\sim${\bf P}$\cdot$({\bf L}$\times${\bf M}) is allowed 
in the energy of the antiferromagnetic-{\bf paraelectric} (AFM-PE) phase, 
where {\textbf L} is the AFM vector. Here, the polarization {\textbf P}   
and the magnetization {\textbf M} are small parameters expanded about
 {\textbf P} = {\textbf M} = 0.
%

Using a symmetry analysis, Ref.~\onlinecite{fennie.prl.08} showed clearly
that such an invariant is allowed in {\bf paraelectric} R$\bar{3}$c magnetic 
A-site ABO$_3$ materials such as FeTiO$_3$ but symmetry forbidden in 
{\bf paraelectric} R$\bar{3}$c magnetic B-site ABO$_3$ materials 
such as the most widely studied multiferroic BiFeO$_3$ (for a more thorough
presentation of this comparison see Ederer and Fennie ~\cite{ederer.fennie}).
Recently, Ref.~\onlinecite{moore.comment} have argued that ``there is
a simple duality between A-site and B-site'' R3c ABO$_3$ materials and
that our Letter~\cite{fennie.prl.08}  ``breaks this duality by ignoring a non-polar
distortion that is directly measured in crystallography \cite{kubel}.''
Here we point out that this is not correct. 
The Comment contains three elementary crystallographic errors:
\begin{enumerate}
\item[(1)] that there is an R$\bar{3}$c reference structure for BiFeO$_3$ with
inversion symmetry on Bi rather than on Fe (an error propagated from
 Kadomtseva et al.~\cite{kad}),
\item[(2)]  that their proposed $\beta$ distortion: (a) generates this structure (b) is
 compatible with the translational symmetry of  BiFeO$_3$,
\item[(3)] that three, not two, fields with different symmetries are
required to obtain the R3c structure of BiFeO$_3$ from the Pm$\bar{3}$m phase,
\end{enumerate}
that make the comment fundamentally wrong and unpublishable.

Let us now address each error in some detail:

(1) {\it R$\bar{3}c$ reference structure}.-- The authors of the Comment claim
that there is another R$\bar{3}$c paraelectric reference structure besides 
the one that we considered in our Letter. This is simply a crystallographic mistake.  
The authors propagate an error present in the work of Ref.~\onlinecite{kad}.
This can be seen by starting, as did Ref.~\onlinecite{moore.comment}, with the
5 atom, cubic perovskite (space group Pm$\bar{3}$m), prototype structure of 
BiFeO$_3$ where Bi sits in the corner of the cube and Fe sits in the 
center of a corner-sharing oxygen octahedra. It should be understood
that the experimental ferroelectric R3c structure of BiFeO$_3$ can be 
viewed as a slightly distorted version of this prototypical structure, e.g., 
Fe, not Bi, is approximately octahedally coordinated as shown in Fig.~\ref{Fig3}. 
One can use standard group theoretic methods to show that there 
is only one distortion (labeled the $\alpha$ distortion in the Comment)
that connects the 5 atom cubic Pm$\bar{3}$m BiFeO$_3$ structure to the 10 atom 
R$\bar{3}$c structure (see Refs.~\onlinecite{stokes.98,stokes.02,stokes} for details). 
This distortion removes 
the inversion center at the 12-fold coordinated site, i.e., the Bi-site, while 
retaining it at the 6-fold coordinated site, i.e., the Fe-site~\cite{ederer.fennie}.
This paraelectric R$\bar{3}$c structure is the one we considered for BiFeO$_3$ 
in our Letter
in which the Fe-site is now at
Wyckoff position 2b and the Bi-site is at Wyckoff position 2a.
The second 10 atom R$\bar{3}$c paraelectric reference structure proposed  
by the authors of the Comment  follows the work of Kadomtseva et al.~\cite{kad}
who mistakingly switched the Bi and Fe positions in the R$\bar{3}$c 
BiFeO$_3$ structure. Subsequently, this results in a structure where Bi, rather 
than Fe, is octahedrally coordinated. This structure is not BiFeO$_3$!

\begin{figure*}[t]
\centering
\includegraphics[width=0.5\textwidth]{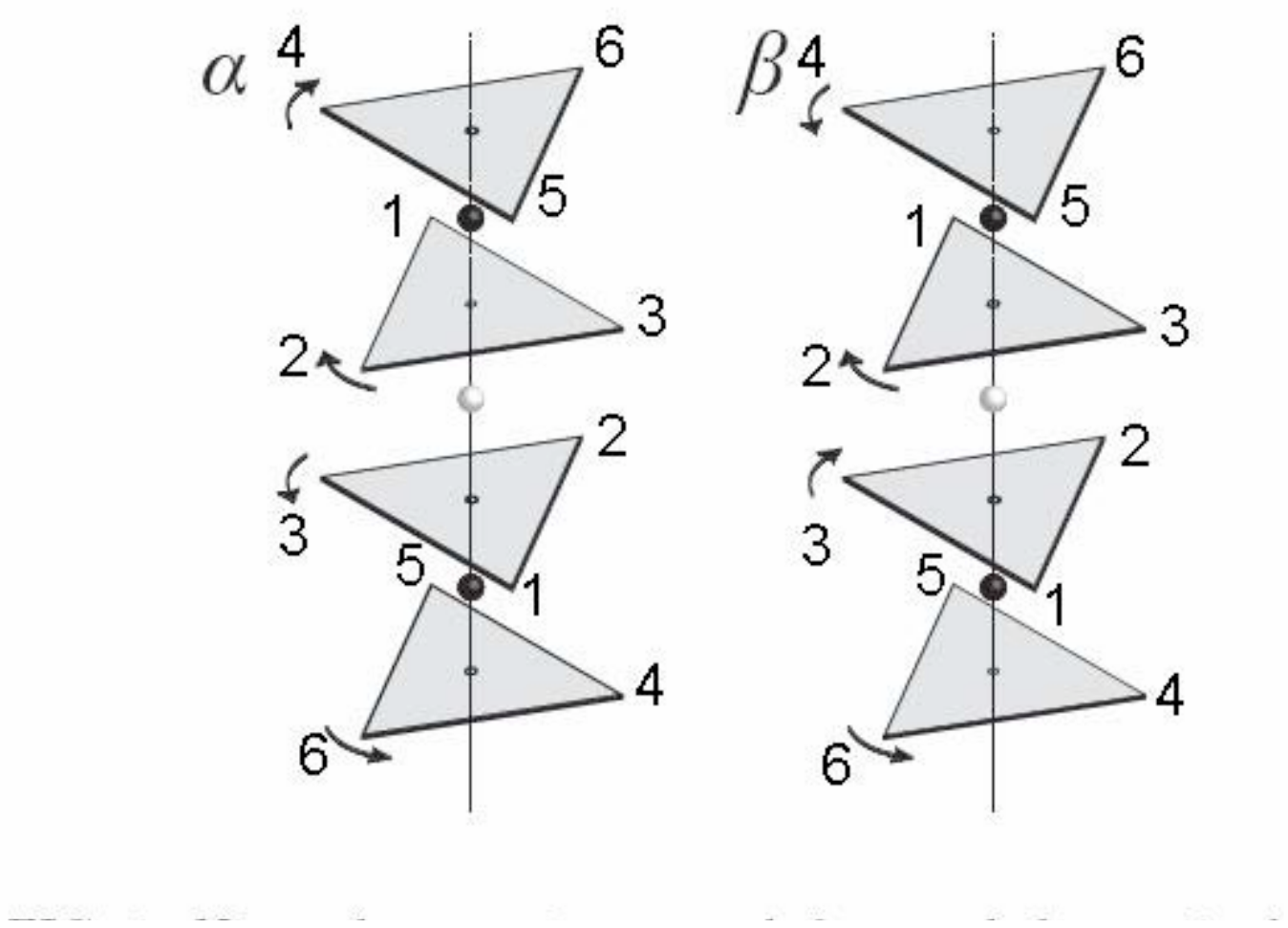}\\
\caption{\label{Fig1} In 10-atom R3c, there are six translationally inequivalent
oxygen atoms, labeled as 1-6. Two atoms both labeled by the same number
are related by a lattice vector translation. The beta distortion proposed
by Ref.~\onlinecite{moore.comment} does not have the correct translational symmetry, 
as each pair of atoms marked by the same number move in opposite directions. 
Note figure taken without permission from Ref.~\onlinecite{moore.comment}
where we added the numbers 
labeling the six unique oxygen atoms in R3c BiFeO$_3$.}
\end{figure*}

\begin{figure*}[t]
\centering
\includegraphics[width=0.5\textwidth]{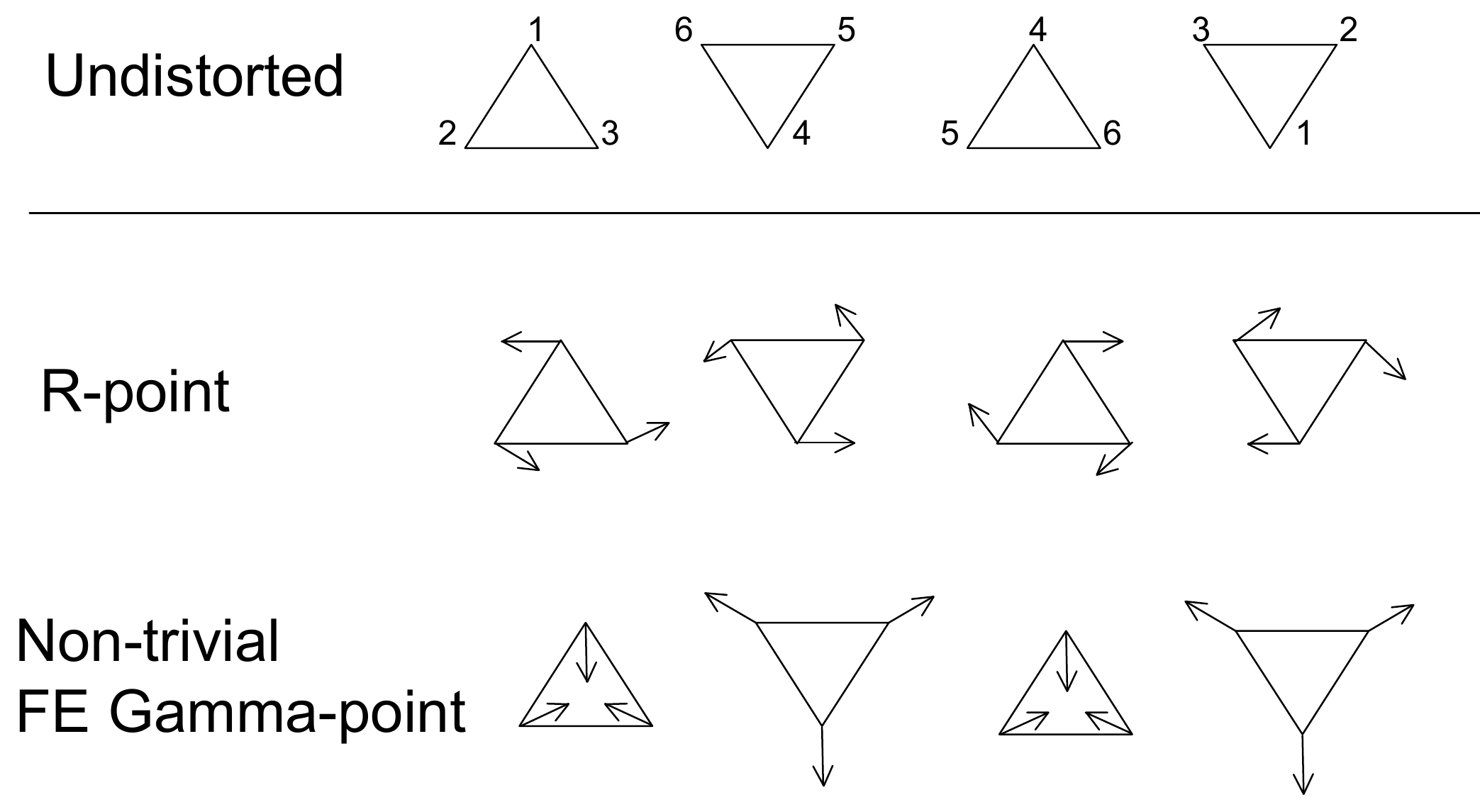}\\
\caption{\label{Fig2} Using only two properly symmetrized lattice-modes of
 the parent Pm$\bar{3}$m  {\it Undistorted} structure, i.e., the R-point distortion
 (the $\alpha$ rotational instability) and the $\Gamma^-_4$ ferroelectric instability, 
 one can account for {\bf all} structural distortions in both the first-principles 
 obtained and the experimental~\cite{kubel} R3c BiFeO$_3$ structures (minus 
 strains) (see Ref.~\onlinecite{stokes.02} for a detailed discussion). 
 Note: here we only show the non-trivial component of the ferroelectric mode (see text
 for a description). }
\end{figure*}

\begin{figure*}[t]
\centering
\includegraphics[width=0.5\textwidth]{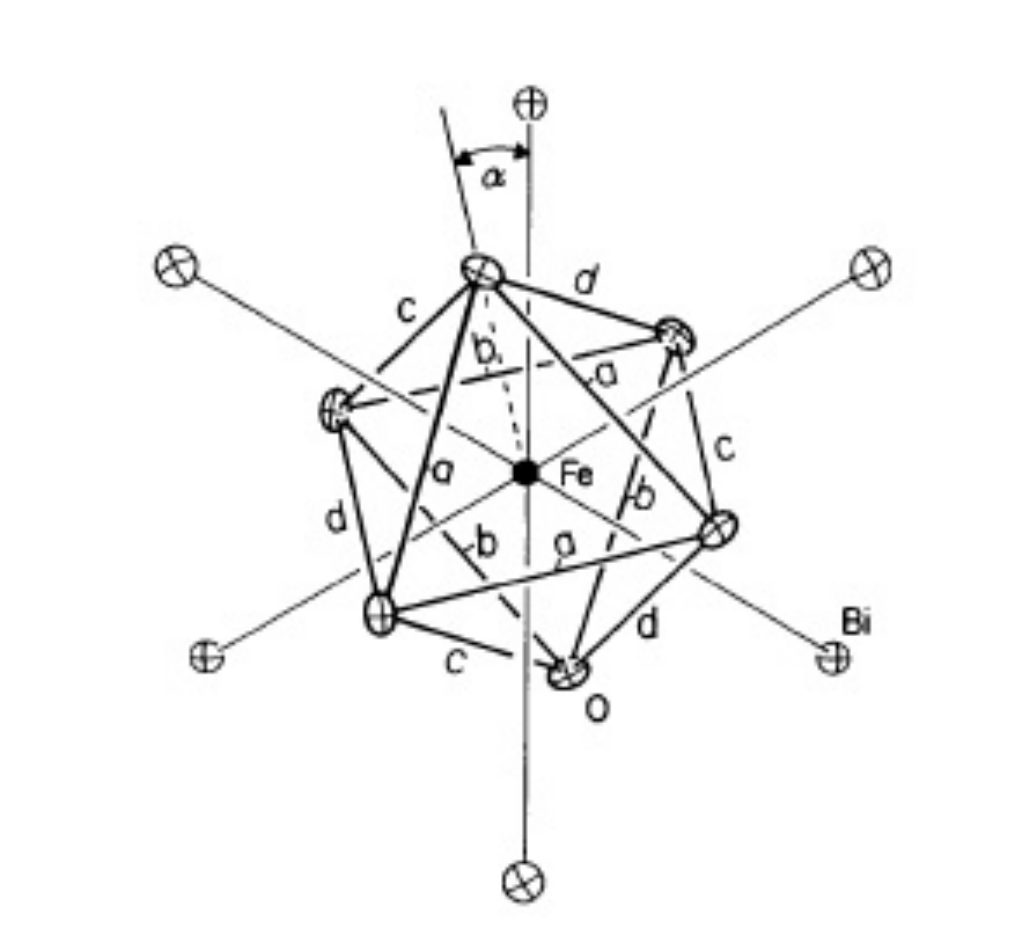}\\
\caption{\label{Fig3} Fe-centered, oxygen  octahedral in ferroelectric R3c BiFeO$_3$. Note figure is copied without permission from Ref.~\onlinecite{kubel}.    }
\end{figure*}

(2) {\it $\beta$ distortion}.-- Since group theory and basic crystallography 
rules out a second  10-atom R$\bar{3}$c reference structure for BiFeO$_3$ 
with inversion symmetry on Bi (point 1), the proposed $\beta$ distortion 
obviously cannot generate it. In fact, the $\beta$ distortion proposed by the
authors of the Comment generates 
a distortion that is incompatible with the translational symmetry of BiFeO$_3$
as we 
now show. In 10-atom R3c BiFeO$_3$ structure, there are six translationally 
inequivalent oxygen atoms, labeled as 1-6 in Figs.~\ref{Fig1} and \ref{Fig2} 
(for example, two atoms both labeled 1 are related by a lattice vector translation). 
As can be clearly seen, the $\beta$ distortion does not have the correct translational 
symmetry, as each pair of atoms marked by the same number move 
in opposite directions. Further analysis shows that the displacement pattern 
generated by $\beta$ has 20 atoms per unit cell and thus does NOT contribute 
to the R$\bar{3}$c (nor the R3c), structure of BiFeO$_3$.

(3) {\it  Decomposition of R3c structure}.-- 
The authors of the Comment claim that three separate distortions of
the 5 atom cubic perovskite structure are required to describe the R3c structure
of BiFeO$_3$ and that my Letter ignores one of them. Although we just 
proved that their proposed third distortion, the $\beta$ distortion, is not
present in BiFeO$_3$ we feel this point is worth an additional response.
To begin with,  first-principles calculations of BiFeO$_3$ in the R3c 
structure~\cite{fennie.prl.08,ederer.prb.06} account for all the distortions present in the experimental 
paper~\cite{kubel} referenced by the authors of the Comment. The claim 
by the authors of the Comment that we ignored distortions present in the 
experimental structure of BiFeO$_3$ is not correct (in fact, it is not clear 
how they came to this conclusion). 
 Additionally, it can easily be shown
that the structure of R3c BiFeO$_3$ is completely described by only two 
fundamental atomic distortions (ignoring strain) of the cubic parent Pm$\bar{3}$m 
phase~\cite{stokes}. Note, the clearest approach to account for the structural distortions 
is to use symmetry-adapted 
lattice modes, i.e., lattice distortions that transform like a particular row of a 
particular irrep. Using these symmetry-adapted modes and other ideas from 
group theory one can show that the distortion from Pm$\bar{3}$m to R3c can be written 
in terms of two different symmetrized lattice modes (not including strains)
(see Ref.~\onlinecite{stokes.02} for a detailed discussion). 
The first mode is a polar  (i.e., q=0) distortion labeled $\Gamma^-_4$. NOTE, 
this is important: this distortion can be thought of as having two parts, a trivial 
polar distortion where the atoms move along +/- [111], and a second one that, 
in the picture of oxygen triangles, causes alternating triangle's to expand and 
contract (as shown Fig.~\ref{Fig2}), but in either case these distortions transform 
like the $\Gamma^-_4$ representation, i.e., a polar vector along [111]. The 
second mode is a non-polar R-point (q= $\pi$/a 111) distortion (see Fig.~\ref{Fig2}), 
which can be thought of as a rotation characterized by a single angle, which 
Ref.~\onlinecite{moore.comment} calls $\alpha$. Using these modes one can 
account for *all* the structural distortions in both the first-principles 
and the experimental~\cite{kubel} R3c BiFeO$_3$ structures (again, minus strains). 
Here is a summary of these two symmterized lattice modes, shown in Fig.~\ref{Fig2},
in the language  of the experimental reference (see Fig.~\ref{Fig3}):
\begin{enumerate}
\item[I.] Starting with 5-atom Pm$\bar{3}$m, add just the R-point instability, 
this doubles the unit cell taking you to 10-atom R$\bar{3}$c with (c=d, a=b). 
\item[II.] Then, starting with R$\bar{3}$c, (for illustrative purposes) 
add only the polar lattice mode 
at Gamma where all the displacements are in a plane perpendicular to [111].
This basis function contains only oxygen by symmetry and is the one I 
referred to as the non-trivial component of the FE mode. This takes you to 
R3c with (c,d,a,b) all different just like in the experiment! 
\item[III.] For another view, go back to Pm$\bar{3}$m and add just the non-trivial FE 
mode. This takes you to R3m (size 1) with c=d, a and b different.
\end{enumerate}
One can understand all this by first realizing that the symmetry adapted 
R-point mode is really a displacement pattern, not a rigid rotation per se.
For example, if we keep in mind the picture of oxygen triangles you can 
visualize the non-trivial FE mode as one triangle getting bigger and the other 
getting small (in Kubel this leads to a $\ne$ b). The R-point *displacement 
pattern*, which translates each oxygen through an equal distance, then effectively 
rotates the smaller triangle through a larger angle than the larger triangle 
(thus breaking c=d).

In summary, the Comment of Ref.~\onlinecite{moore.comment} have argued
that ``there is a simple duality between A-site and B-site'' R3c ABO$_3$ 
materials and that our Letter~\cite{fennie.prl.08}  ``breaks this duality by 
ignoring a non-polar distortion that is directly measured in 
crystallography \cite{kubel}.''
We have shown that because the structure of BiFeO$_3$ is known to be 
a network of corner shared Fe-centered oxygen octahedra, then group 
theoretic analysis and basic crystallography tells us that this is impossible. 
Additionally we have shown that their proposed $\beta$ 
distortion breaks translation symmetry of the known BiFeO$_3$ structure
and therefore can not contribute to the BiFeO$_3$ structure. Finally we have
shown that first-principles calculations account for the distortions present in the
experimental structure and that two, rather than three, unique symmetry modes 
are needed to describe this structure.
The authors 
have not only propagated an error that has appeared in the 
literature~\cite{kad}, but in trying to reconcile this previous work and our
own, have created several new errors as well. Their comment is fundamentally 
wrong and should not be published.

The author acknowledges useful discussions with Claude Ederer, Joel Moore,
Jeffery Neaton, and Karin Rabe.

\end{document}